\documentclass[11pt,preprint]{aastex} 
\usepackage{graphicx}

\newcommand{\GPB}{\mbox{\em{GP-B}}}
\newcommand{\IMP}{\mbox{IM~Peg}}
\newcommand{\uv}{\mbox{$u$-$v$}}

\newcommand{\muas}{\mbox{$\mu$as}}
\newcommand{\muasyr}{\mbox{$\mu$as yr$^{-1}$}}
\newcommand{\masyr}{\mbox{mas~yr$^{-1}$}}

\newcommand{\Msol}{\mbox{$M_\sun$}}
\newcommand{\Rsol}{\mbox{$R_\sun$}}
\newcommand{\RA}{$\alpha$} 
 
\newcommand{\DEC}{$\delta$} 

\newcommand{\Ba}{\objectname[ICRF J225307.3+194234]{B2250+194}}
\newcommand{\Bb}{\objectname[87GB 225231.0+171747]{B2252+172}}
\newcommand{\Ra}[4]{\mbox{${#1}^{\rm h} \; {#2}^{\rm m} \; {#3}\fs{#4} $}}
\newcommand{\dec}[4]{\mbox{${#1}\arcdeg \; {#2}\arcmin \; {#3}\farcs{#4} $}}

\newcommand{\kmsMpc}{\mbox{km s$^{-1}$ Mpc$^{-1}$}}

\newcommand{\Aca}{\mbox{$A_{\rm c\alpha}$}}
\newcommand{\Acd}{\mbox{$A_{\rm c\delta}$}}
\newcommand{\Asa}{\mbox{$A_{\rm s\alpha}$}}
\newcommand{\Asd}{\mbox{$A_{\rm s\delta}$}}

\newcommand{\al}{\mbox{$\alpha$}}
\newcommand{\de}{\mbox{$\delta$}}
\newcommand{\mua}{\mbox{$\mu_{\alpha}$}}
\newcommand{\mud}{\mbox{$\mu_{\delta}$}}

\newcommand{\chinu}{\mbox{$\chi^2_{\nu}$}}

\newcommand{\pone}{\mbox{Paper~I}}
\newcommand{\ptwo}{\mbox{Paper~II}}
\newcommand{\pthree}{\mbox{Paper~III}}
\newcommand{\pfour}{\mbox{Paper~IV}}
\newcommand{\pfive}{\mbox{Paper~V}}
\newcommand{\psix}{\mbox{Paper~VI}}
\newcommand{\pseven}{\mbox{Paper VII}}

\defcitealias{GPB-I}{Paper~I}
\defcitealias{GPB-II}{Paper~II}
\defcitealias{GPB-III}{Paper~III}
\defcitealias{GPB-IV}{Paper~IV}
\defcitealias{GPB-V}{Paper~V}
\defcitealias{GPB-VI}{Paper~VI}
\defcitealias{GPB-VII}{Paper VII}

\begin{document}

\title{VLBI for Gravity Probe B: The Guide Star, IM Pegasi}

\author{N. Bartel\altaffilmark{1}, M. F. Bietenholz\altaffilmark{1, 2},
D. E. Lebach\altaffilmark{3}, R. R. Ransom\altaffilmark{4, 5}, M. I. Ratner\altaffilmark{3}, and
I. I. Shapiro\altaffilmark{3}}

\altaffiltext{1}{Department of Physics and Astronomy, York University,
4700 Keele Street, Toronto, ON M3J 1P3, Canada}

\altaffiltext{2}{Hartebeesthoek Radio Astronomy Observatory,
PO Box 443, Krugersdorp 1740, South Africa}

\altaffiltext{3}{Harvard-Smithsonian Center for Astrophysics, 60
Garden Street, Cambridge, MA 02138, USA}

\altaffiltext{4}{Okanagan College, 583 Duncan Avenue
West, Penticton, B.C., V2A 8E1, Canada} 

\altaffiltext{5}{National
Research Council of Canada, Herzberg Institute of Astrophysics,
Dominion Radio Astrophysical Observatory, P.O. Box 248, Penticton,
B.C., V2A 6K3, Canada}

\begin{abstract} 

We review the radio very long baseline interferometry (VLBI)  observations of the guide star,
\IMP, and three compact extragalactic reference sources,
made in support of the NASA/Stanford gyroscope relativity mission, \GPB. The main goal of the observations was the determination of the proper motion of \IMP\ relative to the distant universe. VLBI observations made between 1997 and 2005 yield a proper motion of \IMP\ of $-20.83$ $\pm $ 0.09 \masyr\ in \al\ and $-27.27$ $\pm $
0.09 \masyr\ in \de\, in a celestial reference frame of extragalactic radio galaxies and quasars virtually identical to the International Celestial Reference Frame 2 (ICRF2).  They also yield a parallax for \IMP\ of 10.37 $\pm$ 0.07~mas, corresponding to a distance of 
96.4 $\pm$ 0.7~pc. The uncertainties are standard errors with statistical and estimated systematic contributions added in quadrature.
These results met the pre-launch requirements of the \GPB\ mission to not discernibly degrade the estimates of the geodetic and 
frame-dragging effects.

\end{abstract}

\keywords{astrometry --- binaries: close ---  gravitation --- 
radio continuum: stars --- radio continuum: galaxies --- relativity ---
stars: activity --- stars: individual (IM~Pegasi) ---
techniques: interferometric}

\section{Introduction}
\label{intro}
Einstein's theory of general relativity (GR) predicts that space-time is curved and warped
by the mass and the angular momentum of a gravitating body. The kinematics of objects in the vicinity of such a body are expected to differ from those based on Newton's theory of gravity. The NASA/Stanford spaceborne mission Gravity Probe~B (\GPB)
was designed to measure these differences with four gyroscopes essentially freely falling around Earth, in a polar orbit at an altitude of about 640 km. According to GR, the gyroscopes' spin axes should precess by $\sim$6\farcs{6}~yr$^{-1}$ in a north-south direction due to the curving of space-time by the Earth's mass, and by $\sim$39~\masyr\ in an eastward direction due to the warping of space-time by the Earth's angular momentum. These precessions are dubbed the ``geodetic" effect and the ``frame-dragging,"  (``Lense-Thirring,'' or ``gravitomagnetic'') effect, respectively, and can be considered as rotations of any near-Earth inertial frame relative to the distant universe. 

The gyroscopes were set spinning within a quartz block with respect to which, through superconducting quantum interference devices (SQUIDs), each of the spin axes' precessions could be measured with an accuracy of 100 \muas\ within $\sim$40 days. The quartz block was bonded to a $\sim$15~cm diameter quartz optical telescope. All of these devices --- the quartz block, the SQUIDs, and the quartz telescope (together termed the probe) --- were placed in a dewar filled with liquid helium and cooled to a temperature of 1.8 K. 
Ideally, the orientation of the probe could be best provided by pointing the telescope
to one suitable quasar, one of the most distant compact types of single object in the universe. However, quasars were too dim for the telescope to track. Only stars in our galaxy were bright enough. The challenge was therefore to find a ``guide star," bright enough for the on-board telescope to track, sufficiently isolated to limit light contamination, and suitably located to not unnecessarily decrease the accuracy of the measurement of the precessions of the gyroscopes. Then the guide star's motion on the sky needed to be measured with respect to quasars or distant galaxies so that the precession of the gyroscopes could be determined with respect to the distant universe.

We described the astronomical effort undertaken with the guide star for the support of the \GPB\ mission, ``VLBI for Gravity Probe B," in seven papers: 
``Overview" \citep[Paper I]{GPB-I}; 
``Monitoring of the structure of the reference sources 3C 454.3, B2250+194, and B2252+172" \citep[Paper II]{GPB-II};
``A limit on the proper motion of the `core' of the quasar 3C 454.3" \citep[Paper III]{GPB-III};
``A new astrometric analysis technique and a comparison with results from other techniques" \citep[Paper IV]{GPB-IV};
``Proper motion and parallax of the guide star, IM Pegasi" \citep[Paper V]{GPB-V};
``The orbit of IM Pegasi and the location of the source of radio emission" \citep[Paper VI]{GPB-VI}; and
``The evolution of the radio structure of IM Pegasi" \citep[Paper VII]{GPB-VII}.  Here we give a synopsis of these papers.

\section{Search for the best guide star and selection of \IMP}
\label{searchsel}

A suitable guide star for \GPB\ had to simultaneously meet a number of requirements:

\begin{itemize}
\item[1.] The motion of the star needed to be known or measurable with an uncertainty small enough not to significantly increase the prelaunch anticipated standard error of \GPB\ of $\le$0.5~\masyr\ in each of the two precessions. The specific requirement was that the standard error be $\leq$0.14~\masyr\ for measurements of the star's motion in each coordinate. In 1990, when we entered the first stage of 
our search for a suitable guide star, no star's (or any other celestial object's) motion on the sky was  known with such accuracy in the optical. Such accuracy had previously been achieved only in the radio. Pioneered by \citet{Shapiro+1979},
astrometric measurements with the radio technique of very long baseline interferometry (VLBI) of a quasar relative to another quasar nearby on the sky yielded a proper-motion estimate with a standard error less than half of the requirement for \GPB\ \citep{MarcaideS1983, Bartel+1986}. Therefore VLBI appeared to be at that time the only option to determine the motion of a suitable guide star with the desired accuracy.

\item[2.] The star needed to be sufficiently bright in the radio, preferably with a flux density at least of order 
1~mJy, for VLBI astrometry measurements to be successful.
 
\item[3.] The star needed to be sufficiently bright in the optical, at least 7th magnitude, for the on-board telescope to detect.

\item[4.] The star's angular distance, $D$, from the ecliptic needed to be between
20$\arcdeg$ and 40$\arcdeg$. The lower limit was set by the requirement that, with a
suitable Sun shield at the telescope, sunlight not enter the dewar and boil off the liquid helium. The upper limit was set by the power requirements of the spacecraft, since the solar panels were fixed to the spacecraft, which would continually point to the guide star, making the power received by the solar panels dependent 
on $D$.

\item[5.]The star needed to be located as close as possible to the celestial equator, 
so as to maximize the sensitivity of the measurement of the frame-dragging precession, which essentially decreases with the cosine of the star's declination. 

\item[6.] The star needed to be sufficiently isolated on the sky. Neighbouring stars or reflection nebulae could possibly intolerably decrease the accuracy of the pointing of the telescope to the star's center of brightness. 
\end{itemize}

To find the most suitable guide star for \GPB, we conducted a VLA survey at 8.4 GHz from 1990 to 1992 of $\sim$1200 stars with V magnitude $\leq$6.0 and a declination between $-20\arcdeg$ and $+40\arcdeg$. No single stars were found in the survey with a flux density above $\sim$1~mJy. In fact, we detected only previously known radio stars, all of them binaries of the RS Canum Venaticorum type \citep[see also, for a later survey,][]{Helfand+1999}.
We identified four potentially suitable guide stars:  $\lambda$~Andromeda \objectname[HR 8961]\ (HR~8961,
$+46\arcdeg$ declination), HR~1099 \objectname[HR 1099]\
($+1\arcdeg$), HR~5110 \objectname[HR 5110]\ ($+37\arcdeg$), and \IMP\
(HR~8703, $+17\arcdeg$). After investigating each of the four candidates with VLBI, studying the sky fields around them in the optical, and checking on possible VLBI reference sources for each of them, we selected \IMP\ as the most suitable for \GPB.

\subsection{Properties of \IMP\  and its surroundings}
\label{prop} 

\IMP\ is a binary with a giant primary and Sun-type secondary. We summarize its optical characteristics and previously known properties in Table \ref{topt}.  The radius of the primary is $\sim$13.3 $\pm$ 0.6 \Rsol, which translates at a distance of \IMP\ of $\sim$100 pc, to an angular radius of $\sim$0.64 $\pm$ 0.03 mas. This radius is comparable to the full-width at half-maximum (FWHM) of the synthesized beam of a global VLBI array operating at 8.4 GHz, and therefore in principle allows for an accurate determination of the location of the radio emission with respect to the center of the primary.

A prerequisite for successful guide star tracking was that the onboard telescope be able to lock sufficiently accurately on the center of the optical disk of the star (or a point with an offset from this center constant during the mission).  One concern here was the appearance of variable dark spots on the primary, shown by Doppler imaging and photometry to cover $\sim$15\% of the star's surface.  
If the variations in the size or location of these spots resulted in a linear trend over the course of the \GPB\ mission in the centroid of optical brightness, relative to the center of the disk of the IM Peg primary, then that trend would have the same effect on the \GPB\ data reduction as an equal error in our VLBI measurement of the proper motion of IM Peg.  Such a trend is not necessarily implausible, given the year-to-year variations in IM Peg's optical brightness and the astrophysical plausibility of variations in typical spot latitudes analogous to the systematic variations of sunspot locations on the Sun. However, \citet{Marsden+2007} showed, through extensive spot mapping spanning the duration of the \GPB\ mission, that any error due to variability in the spot pattern would yield no more than a~0.04~\masyr\ drift in the optical center.

A second concern was the surroundings of IM Peg on the sky.  Any star or nebula within the field of view of the telescope when locked on \IMP, possibly combined with \IMP's photometric variability, could cause systematic astrometric errors.  However, for  \IMP, 
no other star within, e.g. $12\arcmin$, is 
brighter than V magnitude 10, which can be compared with the corresponding relative large brightness of \IMP, that varied during the mission only between V magnitude 5.7 and 6.0. All in all, through extensive optical and even 
millimetric observations of the CO(J=1$-$0) line (the latter to search for any molecular cloud that could be associated with a reflection nebula), we constrained any astrometric systematic error to a negligible value: The combined error from dark spots, neighbouring stars, and a nebula, combined with the photometric variability of \IMP, we estimated to be smaller than 0.05~\masyr. 
 
\begin{deluxetable}{l c c@{~~~}c c c}
\tabletypesize{\scriptsize}
\tablecaption{Optical characteristics and previously known properties of \IMP}
\tablewidth{0pt}
\tablehead{
  \colhead{Parameter} &
  \colhead{} &
  \multicolumn{2}{c}{Value} &
  \colhead{} &
  \colhead{Reference}
}
\startdata

{\em Hipparcos}, $\alpha (J2000)$, epoch 1991.25 & & \multicolumn{2}{c} {\Ra{22}{53}{02}{278706}  $\pm$ 0.63~mas} & &  1 \\
{\em Hipparcos}, $\delta (J2000)$, epoch 1991.25 & &\multicolumn{2}{c} {\dec{16}{50}{28}{53982}  $\pm$ 0.43~mas} & &  1 \\
{\em Hipparcos}, parallax (mas)      &  & \multicolumn{2}{c}{$10.33 \pm 0.76$}     &  &   1 \\
{\em Hipparcos}, distance (pc)        &  & \multicolumn{2}{c}{$96.8^{+7.7}_{-6.2}$} & &   \\
                                      &  &                      &                                   &  & \\
\tableline
\multicolumn{6}{c}{Stellar Properties\tablenotemark{a}} \\
\tableline
                                      &  &                      &                                   &  & \\
V magnitude range during mission     &  &    5.7 to 6.0            &                                                 &  & 2    \\
Mass ($\Msol$)                        &  & $1.8 \pm 0.2$   & $1.0 \pm 0.1$                          &  & 3,3 \\
Spectral Type                         &  & K2~III          & G~V\tablenotemark{b}              &  & 4 \\
$T_{\rm{eff}}$ (K)                    &  & $4550 \pm 50$   & $5650 \pm 200$\tablenotemark{b}        &  & 4,3 \\
Radius ($\Rsol$)                      &  & $13.3 \pm 0.6$  & $1.00 \pm 0.07$\tablenotemark{b}       &  & 4,3 \\
Radius (mas)\tablenotemark{c}         &  & $0.64 \pm 0.03$ & $0.048 \pm 0.004$\tablenotemark{b}     &  & 4,3 \\
                                      &  &                      &                                   &  & \\
\tableline
\multicolumn{6}{c}{Orbital Elements\tablenotemark{a}} \\
\tableline
                                      &  &                      &                                   &  & \\
$a \sin i$ ($\Rsol$)                  &  & $16.70 \pm 0.02$     & $30.34 \pm 0.03$                  &  & 3,3 \\
$a \sin i$ (mas)\tablenotemark{c}     &  & $0.806$              & $1.464$                           &  & \\
$P$ (days)                            &  & \multicolumn{2}{c}{$24.64877 \pm 0.00003$}               &  & 3 \\
$i$ ($\arcdeg$)                       &  & \multicolumn{2}{c}{$65$ to $80$, $>$55}                   &  & 4,5 \\
$e$                                   &  & \multicolumn{2}{c}{$0.0$ (assumed)}                      &  & 4 \\
$T_{\rm conj}$ (HJD)\tablenotemark{d}     &  & \multicolumn{2}{c}{$2450342.905 \pm 0.004$}              &  & 3 \\
\enddata
\tablenotetext{a}{Two entries correspond to the two
stars of the binary system, with entries for the primary listed
first. First reference is for the first entry, second
reference, if present, is for the second entry.}
\tablenotetext{b}{The spectral type, effective temperature, and radius
of the secondary are inferred from the flux ratios (at two
wavelengths) of the two stellar components and the values for the
radius and effective temperature of the primary under the assumption
that the secondary is a main sequence star.}
\tablenotetext{c}{Computed for a system distance of $96.4 \pm 0.7$~pc.
The uncertainty in the $a \sin i$ value in $\Rsol$ units is not
propagated into mas, since the uncertainty in the inclination is the
dominant source of error in any spectroscopic determination of the
semimajor axis.}
\tablenotetext{d}{Heliocentric time of conjunction with the K2~III
primary behind the secondary.}
\tablerefs{
  1. {\em Hipparcos} Catalogue \citep{PerrymanESAshort1997};\phn
  2. \pfive;\phn
  3. \citet{Marsden+2005};\phn
  4. \citet{BerdyuginaIT1999} ($e=0.006\pm0.007$);\phn
  5. \citet{Lebach+1999}.
  }
\label{topt}
\end{deluxetable}

\begin{deluxetable}{l l l l l l l}
\tabletypesize{\scriptsize}
\tablecaption{Characteristics of the sources}
\tablewidth{0pt}
\tablehead{
\colhead{Source} &  \colhead{Type} & \multicolumn{2}{c}{Separation}  & Flux density\tablenotemark{a} &   \colhead{Redshift}
                                    & \colhead{Distance}\tablenotemark{b}\\
                 &                 & \colhead{$\Delta$\al(\arcdeg)}  & \colhead{$\Delta$\de(\arcdeg)} &  \colhead{(Jy)} 
                 &                   & \colhead{(Mpc)} 
}
\startdata
3C~454.3 & quasar         & \nodata           & \nodata & 7  -- 10       & 0.859    & 1610       \\ 
B2250+194 & galaxy        &  $-0.2$           & 3.6     & 0.35   -- 0.45 & 0.28     & \phn880    \\
B2252+172 & unidentified  &     \phn 0.4      & 1.4     & 0.017          & \nodata  & \nodata     \\
\\
IM Peg  & RS CVn          &  $-0.1$           & 0.7     & 0.0002 -- 0.08  & 0.0      & \phn\phn\phn0.0  \\
\enddata
\tablenotetext{a}{The range gives the lowest and highest flux density measured at 8.4 GHz with the VLA during the course of the
observations, 1997 January to 2005 July.}
\tablenotetext{b}{The angular diameter distance for a flat universe with Hubble constant, 
$H_0$=70~\kmsMpc, and normalized density parameters, $\Omega_M=0.27$ and $\Omega_{\lambda}$=0.73 
.} 
\label{t2refsources}
\end{deluxetable}

\section{Radio reference sources for  \IMP\ in the distant universe}
\label{ref}
Our goal was to determine the motion on the sky of \IMP\ relative to the distant universe, with a standard error not exceeding 0.14~\masyr\ in either coordinate. To achieve this goal, we selected radio reference sources with respect to which the motion of \IMP\ could be measured. These sources needed to be at cosmological distances, be sufficiently strong and compact for VLBI observations, and located nearby to \IMP\ on the sky so as to minimize systematic astrometric errors. 

\begin{figure}[htp]
\centering
\centering
\includegraphics[width=\textwidth]{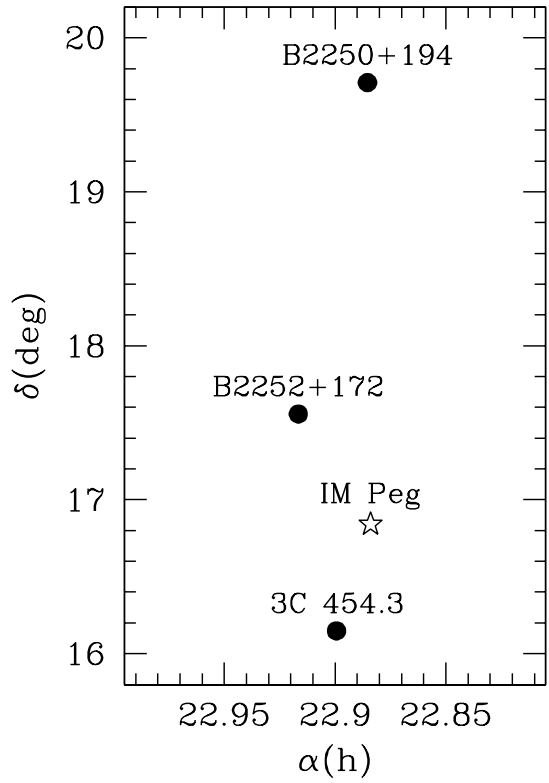} 
\caption{Relative positions (J2000) on the sky of the four radio sources used for \GPB\ astrometry. 
The $\alpha$ (east-west) and $\delta$ (north-south) directions on the plot are on the same scale.
Figure taken from \pthree.
}
\label{f1skypos} 
\end{figure}

\begin{figure}
\centering
\includegraphics[width=0.49\linewidth]{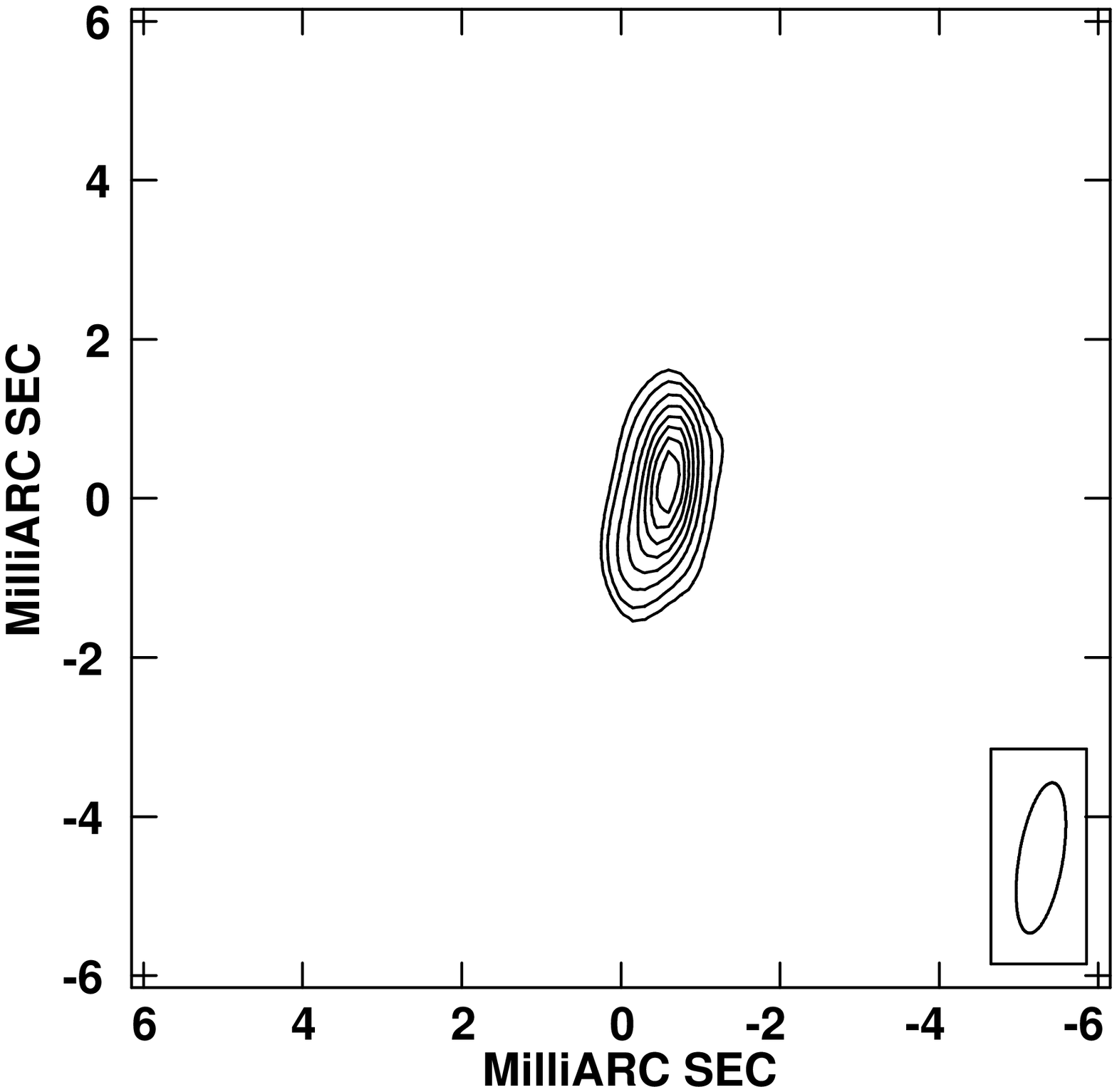}
\includegraphics[width=0.49\linewidth]{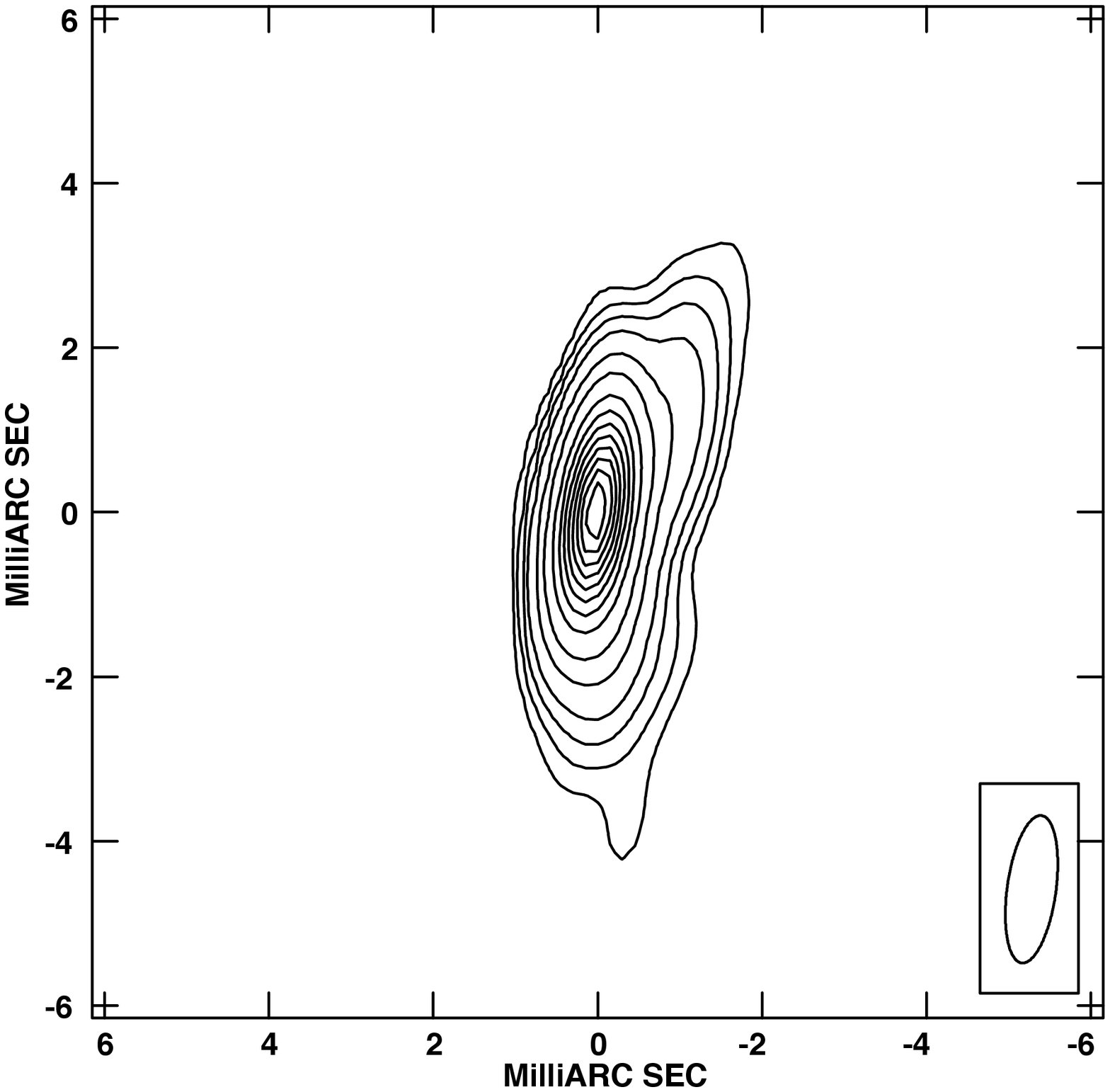}
\includegraphics[width=0.49\linewidth]{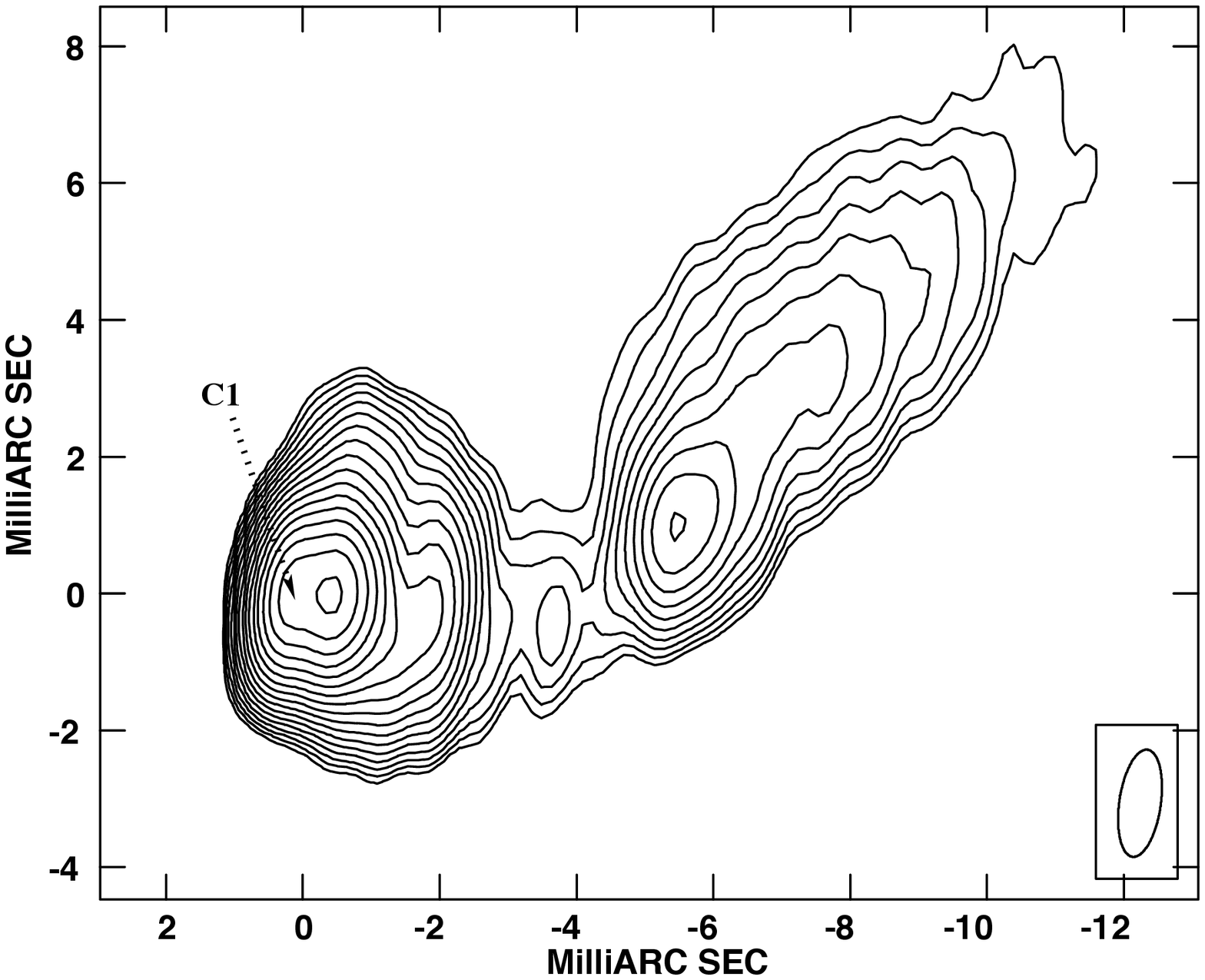}
\includegraphics[width=0.49\linewidth]{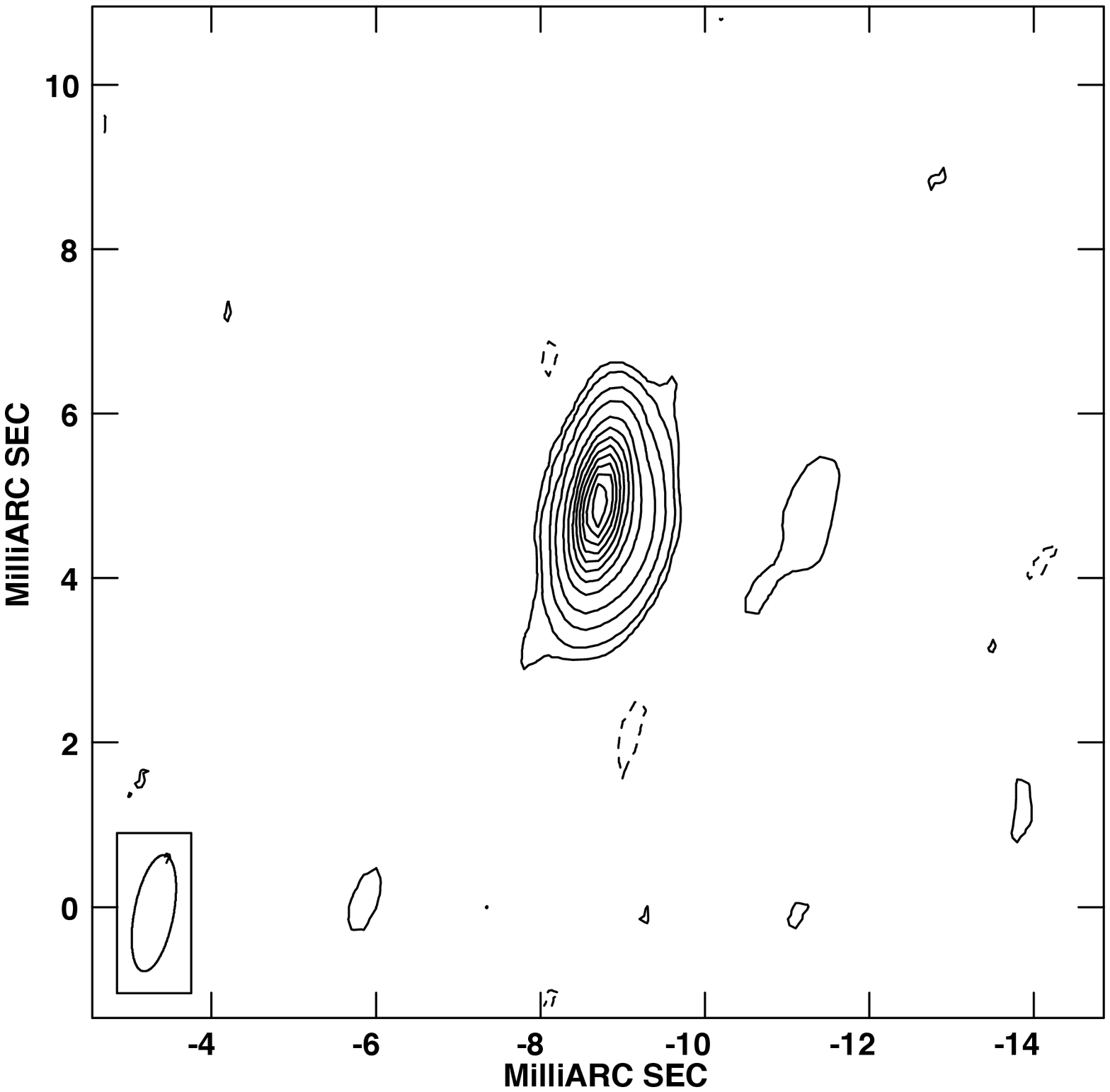}
\caption{
VLBI images at 8.4 GHz of \IMP\ (upper left panel) on 2004 December~11, and the three reference sources, 3C 454.3 (lower left panel) also
on 2004 December~11, B2250+194 (upper right panel) on 2000 November 5, and B2252+172 (lower right panel) on 2003 May 18. The restoring beam characterizing the resolution in the image is given for each panel at its 50$\%$ contour (FWHM) in the inset. North is up and east to the left. For more information, 
see \ptwo\ and \pseven. Figure taken from \pone, but rearranged.}
\label{f2images} 
\end{figure}

We selected three reference sources, with the strong quasar, \objectname{3C454.3}, as the main one. The other two were the active nucleus of the radio galaxy, \objectname[]{B2250+194}, and
the unidentified source (most likely also a quasar or a radio galaxy), \objectname[]{B2252+172}.  We display their positions on the sky, together with that of \IMP, in Figure~\ref{f1skypos}, and list their characteristics in Table~\ref{t2refsources}. 
In Figure~\ref{f2images} we present typical VLBI images at 8.4 GHz of \IMP\ and each of the three references sources.

All the sources are approximately located along the same axis (north-south), allowing perhaps for a more accurate relative correction of tropospheric and ionospheric effects. However, each of the three sources has advantages and disadvantages as reference sources for \IMP. The quasar 3C~454.3 is both the strongest and the closest on the sky to \IMP. It was used as a reference source for \IMP\ from 1991 to 1994 \citep{Lestrade+1999}, and could therefore extend the time baseline of our VLBI observations. The disadvantage is the complexity of 3C~454.3's structure and its superluminal nature. In Figure~\ref{f2images} we identified six components of which the easternmost component, C1, is still compact at the highest angular resolution, has a flat or inverted high-frequency spectrum \citep{Pagels+2004}, and is
located almost exactly at the same position (\ptwo) as the core of the quasar seen at 43 GHz \citep{Jorstad+2001, Jorstad+2005}. The component C1 was therefore considered to be the core at 8.4 GHz, the part most closely related to the supermassive black hole and to the center of mass of the quasar; it was thus the reference
point in the quasar's structure for our astrometric measurements. The other components are moving away from C1, with apparent speeds of up to $\sim$5~c  (\ptwo).  
The remaining sources were both rather compact, so that the brightness peak could simply be taken as the reference point in the structure; their disadvantages were that B2250+194 was located relatively far away from \IMP\ and B2252+172 was rather faint (with an unknown redshift). 

In addition, we used geodetic and astrometric VLBI observations regularly conducted over many years to determine the limit on the proper motion of two of the three reference sources, 3C~454.3 and B2250+194. These determinations were made without consideration of a particular reference point in the structure of the two sources but were made with respect to a celestial reference frame, CRF. This frame is defined by a large number of extragalactic radio sources distributed over the sky and is for our purposes essentially the same as the International Celestial Reference Frame, ICRF2 \citep{FeyGJ2009}. 

All in all, we took C1 in 3C~454.3 as the primary reference point with respect to which the motion of \IMP\ was measured. The other measurements served to determine the stability of C1 on the sky so that finally the motion of \IMP\ could be determined with respect to the distant universe.

\section{Astrometric VLBI observations of \IMP\ and its radio reference sources, and data analysis}
\label{vlbiobsanalysis}

To determine the motion of \IMP\ relative to C1 and test C1's stability relative to the brightness peaks of the other two reference sources, we took into account: 

\begin{itemize}
\item[1.] \IMP\ is catalogued as a binary. However, we could not rule out that it is perhaps part of an extended triple or even larger multiple system. Therefore, to ensure our obtaining a sufficiently accurate value of the proper motion of \IMP\ for the $\sim$15 months \GPB\ data-collection interval in 2004 and 2005, we spread a sufficiently large number of VLBI measurements over eight and a half years to be able to determine any nearly constant change of the proper motion with time (``proper acceleration"), or place a limit on it.
\item[2.] For the best determination of \IMP's parallax, we spread the VLBI measurement epochs appropriately to cover the different seasons each year. Similarly, to allow us to model any orbital component of the motion of the radio emission, we carefully distributed the epochs over the phase of the binary orbit, which could be computed from earlier spectroscopic results.
\item[3.] To measure  the time-variable brightness distributions of \IMP\ and the three reference sources, and in particular to identify C1 in 3C 454.3, we selected each VLBI observation session to be long enough to ensure that the \uv\ coverage was sufficiently dense for us to obtain excellent imaging of the sources.
\item[4.] Since we expected \IMP\ to fluctuate strongly in radio brightness, and since we knew that radio emission from B2252+174 was relatively weak, we needed our VLBI array to be sufficiently sensitive to nearly guarantee detection of the sources in each observing session.
\item[5.] Lastly, we needed an observing frequency for which receivers were available at each of the telescopes of the VLBI array, and which guaranteed sufficiently high angular resolution and low system temperatures.
\end{itemize}

All said, 35 sessions of astrometric VLBI observations were made between 1997 January 16 and 2005 July 16 with about four sessions every year, each about 11 to 15 h long. The observations were made with a global array of 12 to 16 radio telescopes
comprised of most or all of: MPIfR's 100 m
telescope at Effelsberg, Germany; NASA/Caltech/JPL's 70 m DSN
telescopes at Robledo, Spain, Goldstone, CA, and Tidbinbilla,
Australia; NRAO's ten 25 m telescopes of the VLBA, across the U.S.A;
NRAO's phased VLA, equivalent to a 130 m telescope, near Socorro, NM;
and, at early times, NRCan's 46 m Algonquin Radio Telescope near
Pembroke, ONT, Canada, and NRAO's 43~m telescope in Green Bank,
WV.  The observing frequency was 8.4 GHz. 

All of these VLBI observations were made to extract interferometric phase delays related to the difference of arrival times of a radio wave from a celestial source at each pair of antennas of a VLBI array. Such observations allow the most accurate astrometric measurements \citep{Shapiro+1979} and can yield relative positions and proper motions referenced to particular points in the structure of the sources with uncertainties as low as $\sim$10~\muas~ and $\sim$10~\muasyr, respectively (for the earliest such measurements, see, e.g., Marcaide \& Shapiro 1983;
Bartel et al. 1986; and for a recent review, see Reid \& Honma 2014). 

The phase delay is given by
\begin{equation}
{\phi(\omega, t) \over \omega},
\end{equation}
where $\phi(\omega, t)$ is the interferometric phase or ``phase" at observing angular frequency, 
$\omega$, and time, $t$. The phase can be expressed as
\begin{equation}
\phi(\omega, t) = \omega \left[ \tau_{\rm{geom}}(t)
 + \tau_{\rm{inst}}(\omega, t)
 + \tau_{\rm{atm}}(\omega, t)
 + \tau_{\rm{struc}}(\omega, t) + \tau_{\rm{noise}}(\omega, t) \right]
 + 2\pi N(\omega, t),
\label{phi}   
\end{equation}
where 
$\tau_{\rm{geom}}(t)$ is the ``geometric delay,'' the difference
in the arrival times of the radio wave in vacuum at
the two antennas;
$\tau_{\rm{inst}}(\omega, t)$ describes the difference in the
instrumental delays (including clock behavior) at the two antenna sites;
$\tau_{\rm{atm}}(\omega, t)$ gives the
difference in radio wave propagation times to the two antennas
due to all tropospheric and ionospheric effects;
$\tau_{\rm{struc}}(\omega, t)$ is the delay contribution from source
structure to account for any non-pointlike brightness distribution of the
celestial source;
$\tau_{\rm{noise}}(\omega, t)$ is the (thermal) noise contribution 
to the phase measurement;
and $N(\omega, t)$ describes the integer number of 2$\pi$ ambiguities, or
``phase wraps,'' in the measurement.

The astrometric information of a VLBI measurement is given by the geometric delay which, apart from
relativistic contributions and those due to Earth's motion relative to the solar system barycenter,
is given by
\begin{equation}
\tau_{\rm{geom}}(t) = {1 \over c}\left[ {\mathbf B}(t) \cdot {\mathbf{\hat s}}(t)\right],
\label{geom} 
\end{equation}
where
$c$ is the speed of light in vacuum,
${\mathbf B}(t)$ is the 3-dimensional vector between two antennas of a
baseline of a VLBI array, and
${\mathbf{\hat s}}(t)$ is the unit vector in the direction of the
observed source.

For phase-delay VLBI observations to succeed, to be able to remove the 2$\pi$ phase ambiguities, and to reduce many sources of astrometric errors, we needed to switch the antennas rapidly between sources. For the first 23 sessions, we used a typical sequence of
3C~454.3 (80~s) - IM~Peg (170~s) - B2250+194 (80~s). For the remaining 12 sessions, B2252+172 was included in the observations and the sequence was altered to 3C~454.3 (80~s) - IM~
Peg (125~s) - B2250+194 (80~s) - 3C~454.3 (80~s) - IM~Peg (125~s) - B2250+194
(80~s) - B2252+172 (90~s).  In addition to the astrometric 8.4 GHz observations, we also observed once 
at 5.0~GHz and once at 15.0~GHz, to allow investigation of the compactness and spectral properties of the components of the sources. Almost all observations were recorded in both right and left circular polarization, and all were processed with the VLBA correlator at Socorro, NM.

We used a custom software package (\pfour) that allowed us to analyze the data efficiently. First, we connected the phases and removed the 2$\pi$ phase ambiguities to generate the correct phase delays. Then we
corrected the phase delays for the effects of the troposphere and the ionosphere with \textit{a priori} estimates. We further processed them with a Kalman filter to model the residuals of the troposphere and the ionosphere, and the clock offset at each telescope from the one chosen as the reference clock for the VLBI array. Then, for the more extended source 
3C~454.3, the structure effects were removed from the phase delays based on the reference point, C1, defined with images made with NRAO's AIPS software package. The other two reference sources were sufficiently compact that no correction for their source structure was needed. 

The relative weakness of \IMP's radio emission, and sometimes also that of B2252+172, made it necessary for us to develop
a special technique for the analysis of \GPB\ VLBI data. It combines the advantages of parametric model-fitting via weighted least-squares applicable for the relatively strong sources 3C 454.3 and B2250+194 with the sensitivity of phase-referenced mapping, the latter required for \IMP\ and B2252+172. The merged analysis technique yielded superior results to the phase-referenced mapping technique (\pfour). 

The radio emission from \IMP\ varied from session to session, reflecting the complex astrophysical 
nature of the environment of the star. In most cases the 
images of \IMP\ showed only one clearly defined component. In these cases the coordinates were determined for the maximum of 
a two-dimensional Gaussian fit to that component. In the remaining cases, there was more than one local maximum of approximately equal brightness in the emission region: in
eight cases there were two such maxima, and in one case there were three, all of them significant.
In such cases, we took the position of IM Peg to be the mean position of the different maxima. 

The final result of the analysis of the \GPB\ VLBI observations was a set of coordinates with statistical standard errors for each source and each observing epoch. These sets were the basis for the study of the motion, or upper bounds on it, of the sources with respect to each other.

In addition, we analyzed geodetic astrometric VLBI observations and made use of interferometric group delays. 
Group delays are given by
\begin{equation}
{d\phi(\omega, t) \over d\omega}.
\end{equation}
These are observations of thousands of sources from all over the sky. These observations are done on a routine basis, independent of the \GPB\ program. For 3C~454.3 we used all available data from a total of 1,119 observing sessions, from 1980 to 2008. 
In support of \GPB, the second reference source, B2250+194, was
included in 38 sessions of routine geodetic astrometric
group-delay observations between 1997 and 2008.
Group-delay observations have the advantage that 2$\pi$ ambiguities are not inherent in the data and that the positions are determined relative to a celestial reference frame (CRF). The disadvantage, however, is that the parameter, $\tau_{\rm{struc}}(\omega, t)$, was not determined and that therefore the exact reference point in the brightness distribution of the sources for the position measurement remained undefined. 

All available geodetic astrometric data with a total of 6.5 million group-delay determinations were processed via least-squares and yielded estimates of coordinates with statistical standard errors for 3C 454.3 and B2250+194 for each observing session, keeping the coordinates of the other sources constant. This set of solutions forms our CRF, 
which for \GPB's purposes is virtually identical \citep[see, e.g.,][]{Petrov+2009} to the ICRF2 \citep{FeyGJ2009}, the most fundamental such frame presently in use.

\section{Positional stability of \IMP's reference sources}
\label{stab}
\begin{figure}[tp]
\centering
\includegraphics[width=0.80\textwidth]{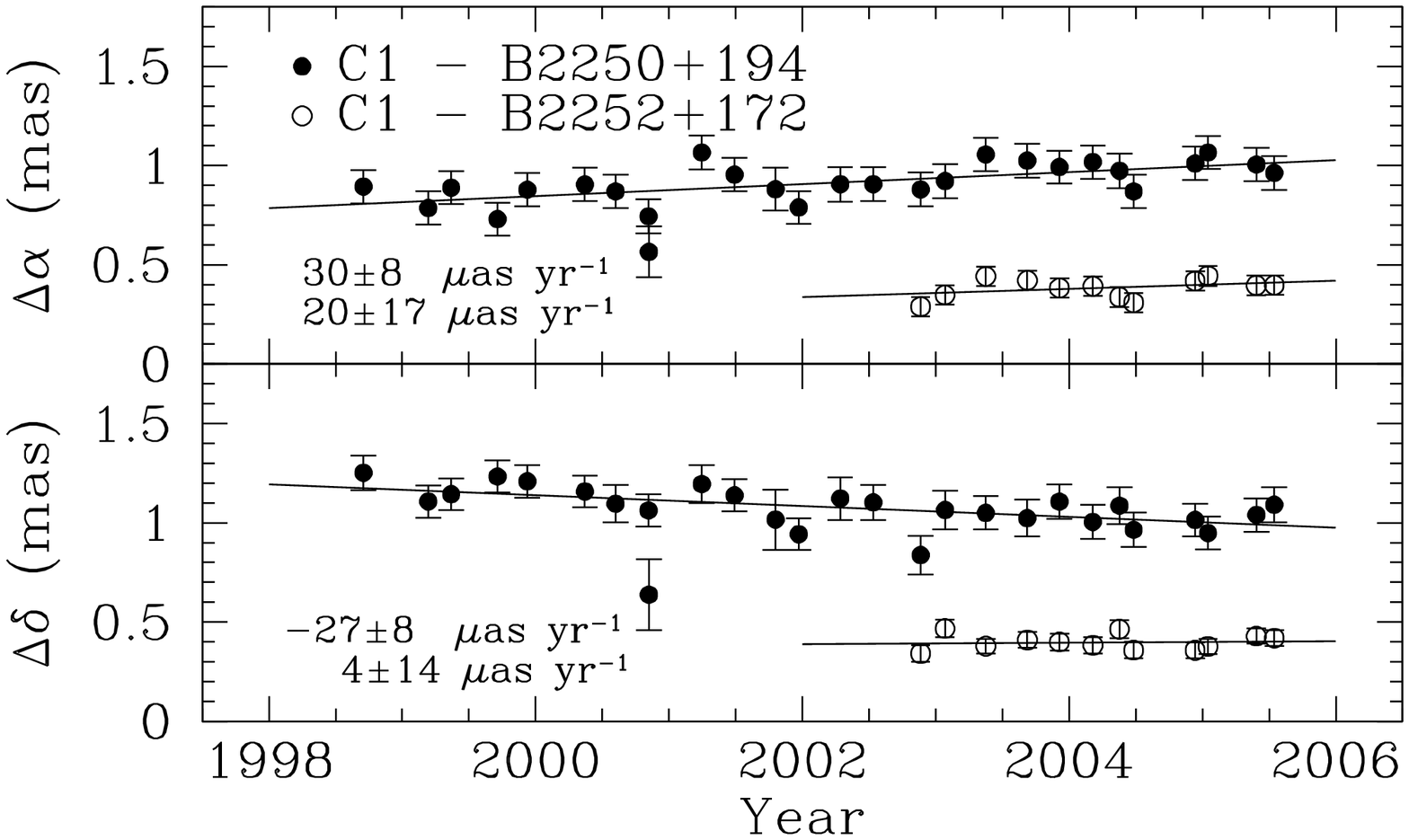}
\includegraphics[width=0.72\textwidth]{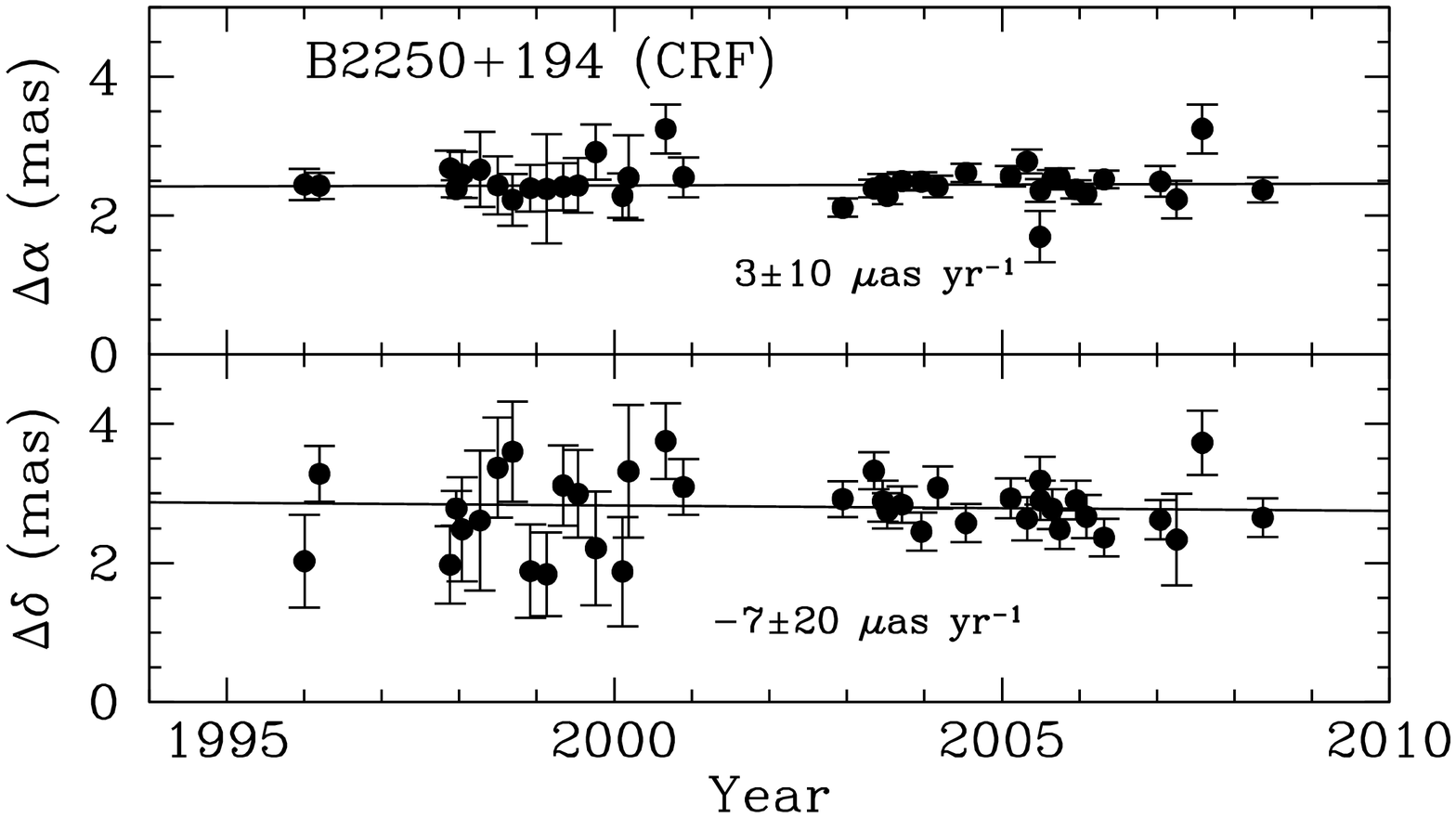}
\caption{Top panel: The estimated coordinates of C1, the core in 3C~454.3, relative to those of
the brightness peaks in B2250+194 and B2252+172 (except for offsets) as a function of
time from \GPB\ VLBI phase-delay observations. The lines give the fit to these coordinates. 
The values for the slopes of the lines and the statistical standard errors
for $\chinu=1$ are also given. Bottom panel: The relative coordinates of B2250+194
as determined from routine geodetic group-delay VLBI observations of up to $\sim$4000 extragalactic sources 
scattered over the sky. Figure taken from \pthree.
}
\label{f3posref}
\end{figure}

The phase-delay determinations of the position of C1, the core of 3C 454.3, relative to those of the brightness peaks of B2250+194 and B2252+172, are plotted in Figure~\ref{f3posref}. 
The group-delay determinations of the position of B2250+194 relative to the CRF are also plotted in
Figure~\ref{f3posref}. See \pthree\ for a similar plot of the group-delay determinations of the position of 3C 454.3 relative to the CRF.

Weighted least-squares straight line fits to each set of position determinations yielded the position at epoch and proper motion for each coordinate separately, along with their statistical standard errors and correlations. In each case the fits to the data gave $\chinu$ ($\chi^2$ per degree of freedom) larger than unity,
indicating systematic errors in the position determinations. These errors can for instance be due to uncorrected effects from source structure and/or  atmospheric and ionospheric variations, and are difficult to quantify. Therefore the total standard errors of the position determinations were determined partly empirically by including a particular constant for each fit. This constant was added in quadrature to the statistical standard errors for each set of position determinations and for each coordinate separately so as to obtain $\chinu=1$ for each fit. 

The resultant proper-motion values with their standard errors are listed for each set of position determinations in the panels of Figure~\ref{f3posref}. 
The  upper panel of Figure~\ref{f3posref} indicates that there appears to be some motion of C1 relative to 
B2250+194 at the 3.5$\sigma$ significance level. The source 3C 454.3 has a long jet, indicating strong activity. 
Some motion of a component like C1 even if located close to the black hole could therefore be expected. 
No significant motion is discernible for B2250+194 relative to B2252+172.
No matter whether there is some motion due to activity near the black hole or not, the 1$\sigma$ upper limits of the proper-motion values for C1 relative to
B2250+194 and B2250+194 relative to the CRF are each smaller than 40 \muasyr. 
Combining the phase-delay proper-motion determination of C1 relative to B2250+194 with the 
group-delay proper-motion determination of B2250+194 relative to the CRF gives for the time period from 1998 to 2005 a 1$\sigma$ standard error for the mean position of C1 in the CRF of 45 and 68 \muas.  The proper motion of C1 relative to the CRF for that same period is $33\pm13$~\muasyr\ and $-35\pm21$~\muasyr\ in \al\ and \de, respectively, which we do not consider significant. We therefore give a 1$\sigma$ upper limit on the magnitude of the proper-motion components of 46 and 56~\muasyr, respectively. The latter pair of determinations is our limit on the level
of stationarity of  the reference point C1 in 3C 454.3 relative to the distant universe (\pthree).

\section{Motions of \IMP\ relative to the distant universe}
\label{motions}
With the stationarity of the reference point in 3C 454.3 determined, the motion of \IMP\ relative to the distant universe could be analyzed. Five kinds of motion needed to be considered: 
\begin{itemize}
\item[1.] proper motion mainly due to the motion of \IMP\ relative to the motion of the Sun in the Galaxy, 
\item[2.] annual parallax due to the orbital motion of Earth around the Sun, together with the much smaller contribution of the motion of the Sun with respect to the solar system barycenter,
\item[3.] change of proper motion over time due to a hypothetical third member of the \IMP\ binary system, 
\item[4.] orbital motion around the common center of mass of the binary, and 
\item[5.] erratic motion of the center of radio emission within the binary system. 
\end{itemize}

\subsection{Astrometric solutions}
To determine the proper motion, parallax, proper acceleration, and orbital motion, we used weighted 
least-squares to fit a linearized model to the 35 determinations of the positions of the reference point in \IMP\ relative to the position of C1 in 3C 454.3. The uncertainties of these position determinations are hard to compute theoretically. We therefore took them from an empirical analysis based on the computation of the root-mean-square (rms) scatter separately in $\alpha$ and $\delta$, of the postfit residuals obtained for the 12 positions of \Bb\ relative to C1 and the 35 positions of \Ba\ also relative to C1. The resulting uncertainty was 0.06 mas in each coordinate. In addition there is the previously discussed uncertainty of the position of C1 and the upper limit of the proper motion of C1 in the CRF that we considered for the standard errors of the position determinations of \IMP\ relative to the CRF. 

The parameters of the model fit were the position of \IMP\ at epoch, proper motion, parallax, and four scalar parameters for the description of the projection on the sky of an orbit with zero-eccentricity \citep{BerdyuginaIT1999} and a period known from optical spectroscopic observations \citep{Marsden+2005}. The fit resulted in rms values of the residuals somewhat different for each coordinate but much larger than the standard errors of the position determinations of \IMP. The source of this large scatter is the astrophysical nature of the radio emission, characterized by motion related to radio brightness changes \citep{Lebach+1999} and plausibly in part to stellar magnetic field changes implied by spot maps from optical spectroscopy \citep[e.g.,][]{Berdyugina+2000}. For the fit we used uniform weighting independently for \al\ and \de, and allowed their errors to have
non-zero correlation. After iteration, we obtained the final set of parameter estimates with statistical standard errors and 
$\chinu=1$. This set is presented in Table~\ref{tfinal}.

For the determination of the proper acceleration, we enlarged the set of positions of 35 epochs by four additional positions of \IMP\  obtained at epochs between 1991 and 1994, also at 8.4 GHz, by \citet{Lestrade+1995} in support of the {\em Hipparcos} mission. These observations considerably extended our time baseline; however, they were only of limited use for the estimate of parameters other than the proper acceleration because of inferior \uv\ coverage, lower angular resolution, and the position determinations not being referred to C1. We found no significant proper acceleration; i.e., none larger than 
one statistical standard error; and so we include no such proper acceleration in the fit used to obtain our final results in Table~\ref{tfinal}.

\begin{figure}
\centering
\includegraphics{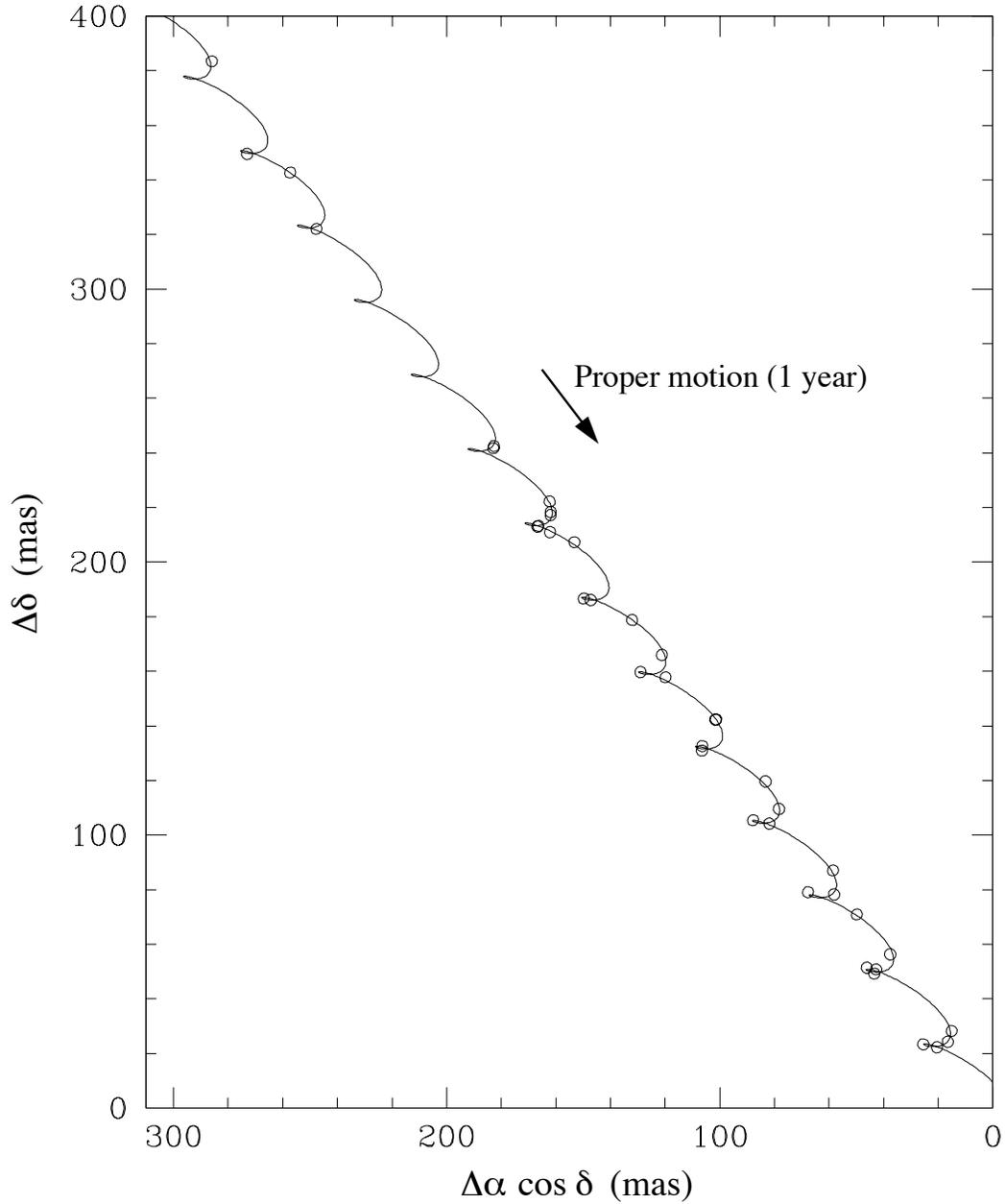}
\caption{
The 35 position estimates of \IMP\ relative to C1. The coordinate origin is arbitrary. The $2\sigma$ errors (here and hereafter in the figure captions in the sense of the full length of $\pm1\sigma$ error bars) are much smaller
than the symbols. The curve shows the fit of the solution, with orbital terms omitted for clarity. 
In addition, the four positions from 1991 to 1994 
are shown (upper left); however, they were not used in the fit. The arrow indicates the direction and its length the magnitude of the proper motion in one year. Figure taken from \pfive.
}  
\label{f4proppar} 
\end{figure}

\begin{figure}
\centering
\includegraphics[width=0.7\textwidth]{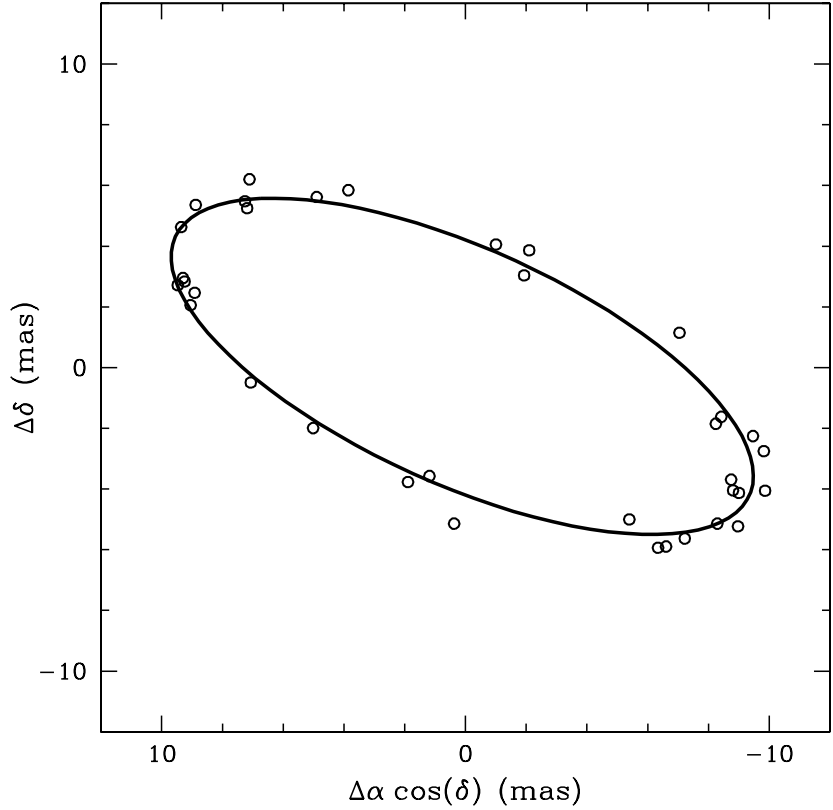}
\caption{
The parallax ellipse (solid line) derived from the nine-parameter fit to the set of 35 position determinations of \IMP. The open circles
give the observed position for each epoch after subtraction of the estimated position at the reference epoch, proper motion, and orbital motion. 
The $2\sigma$ errors (of position determinations relative to C1) are about a third of the size of the circles. The relatively large scatter is most likely 
dominated by the fluctuations of the positions of the stellar radio emission relative to the center of the disk of the \IMP\ primary.
The model positions (not shown) are almost exactly on the plotted parallax ellipse and close to the observed positions.
The parallax ellipse, however, is slightly time-dependent due to the influence of planets. Over the course of our observations of several years, the parallax ellipse in the figure shifts by about the width of the line. 
}
\label{f5paraellipse}
\end{figure}

\begin{figure}
\centering
\includegraphics[width=0.7\textwidth]{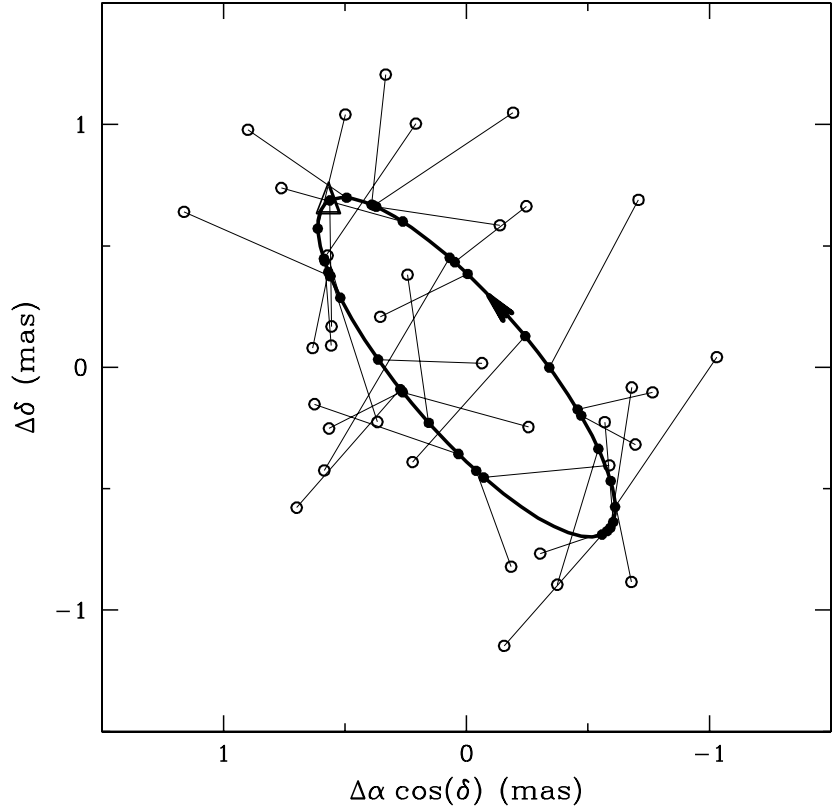}
\caption{The orbit (solid line) of the primary of \IMP\ derived
from the nine-parameter fit to the set of 35 position determinations of \IMP\ with the arrow giving the direction of motion.
The large triangle on the upper left segment of the ellipse marks the ascending node of the inferred orbit, under the convention that at an ascending node an orbiting stellar component recedes from Earth. Thus the part of the ellipse with the arrow on it is closest to Earth.
The open circles give the observed position for each epoch after subtraction of the estimated
position at the reference epoch, proper motion, and parallax.   The $2\sigma$ errors (for the position determinations relative to C1) are about twice the size of the circles. A solid line connects each observed position
with the corresponding fitted position indicated by a dot on the estimated
orbit. The large scatter of the observed positions about the model orbit ellipse is again most likely dominated 
by the fluctuations of the positions of the stellar radio emission relative to the center of the disk of the \IMP\ primary.
Figure adapted from \psix.
}
\label{f6Morbit}
\end{figure}

\begin{figure}[tp]
\centering
\includegraphics[height=5.9in,trim=0 0.0in 0 1.1in,clip]
{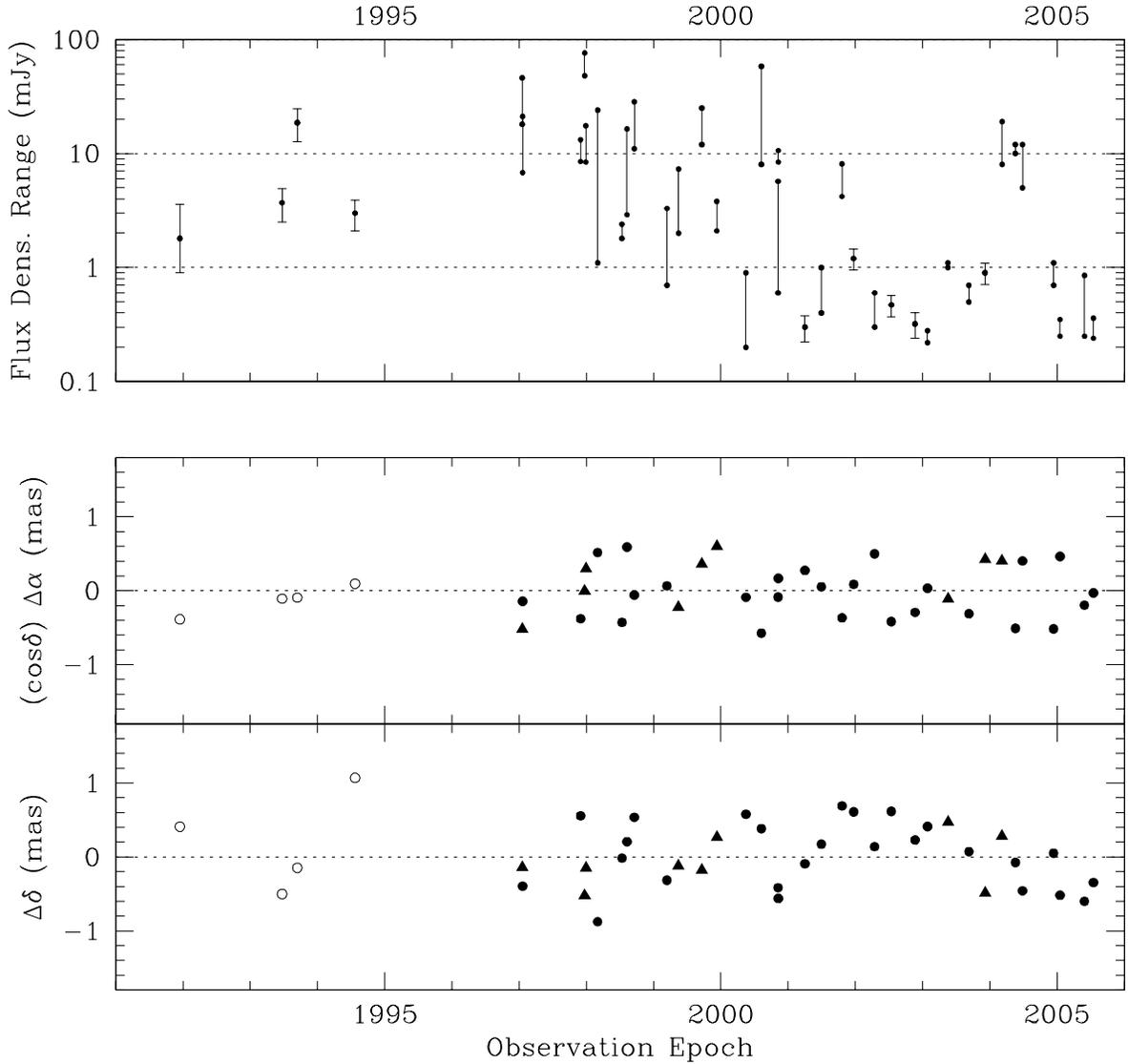}
\caption{
The upper panel shows flux densities for our 39 VLBI sessions, either as a range of values measured with the VLA within a session or as a single value with 1$\sigma$ error bars obtained mostly from the VLBI images. The two lower panels show 
the position residuals for all sessions.  We plot
unweighted points as open circles and weighted points as either closed circles
or triangles, with each triangle indicating a position computed as the mean
position of the two or three resolved peaks in the stellar radio image for
that session. The $2\sigma$ errors (for position determinations relative to C1) are about the size of the symbols.
For more information, see \pfive, from which this figure was taken.
}  
\label{f7resvt} \end{figure}

\begin{figure}
\centering
\includegraphics[width=0.95\textwidth]{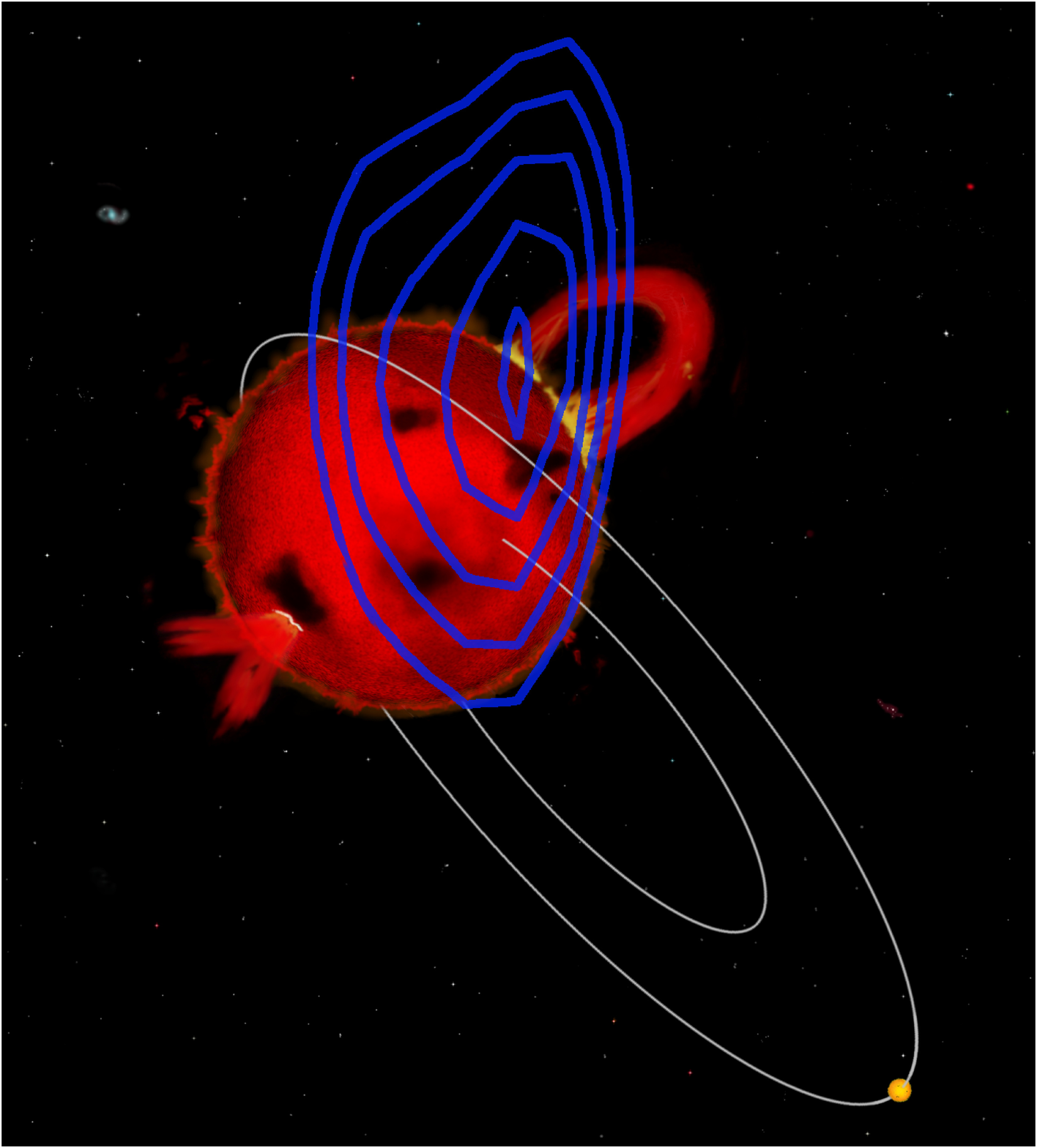}
\caption{Artist's three-dimensional rendering of the \IMP\ binary
system as seen from Earth. The
primary is the larger red star with dark spots, with the spin axis parallel to the orbit normal, 
and mass ejection along magnetic field lines in the polar region that could give rise to the observed radio emission.
The blue contours show, as an example, the radio
brightness distribution as observed on 2003 May 18 relative to the primary (see \pseven). Similar radio emission
relative to the primary at approximately equal phases in its orbit were observed on 1997 November 30 and 2000 August 7 as shown in Figure~\ref{f9image}. The secondary is
the smaller yellow star.  The projected orbit of the primary is the
same as the inferred radio source orbit shown in Figure~\ref{f6Morbit}.
The projected orbit of the secondary and the diameters of the two stars
correspond to the nominal values given in Table~\ref{topt}. The system is shown with the primary at its
ascending node.  Note that the primary's visible pole is near its southeast side. The size of the figure is approximately 3.8~mas by 3.4~mas.
}
\label{f8IMschemradio}
\end{figure}

Figure~\ref{f4proppar} displays the 39 position determinations and the fit to the 35 position determinations, with the four earliest positions not used in the fit.
The effect of the proper motion and the parallax can be clearly seen in the data as well as in the fit.
Figure~\ref{f5paraellipse} displays the effect of the parallax still more clearly.  It shows the 35 position determinations with the estimated position at epoch, proper motion, and the primary's orbital motion subtracted. It also shows the estimated parallax ellipse. 
Figure~\ref{f6Morbit} displays the orbit clearly. Again we plot the 35 position determinations with the estimated position at epoch, proper motion, but now with parallax rather than orbital motion subtracted. Also shown are 
the estimated orbit and the corresponding positions predicted from the orbit model. Figure~\ref{f7resvt} shows the 39 position determinations and the fit to the above 35 positions, but now with all model contributions, namely the estimated position at epoch, proper motion, parallax, and the primary's orbital motion subtracted. We compare them to the corresponding flux densities of \IMP. The scatter of the position residuals appears to be random with certainly no correlation to the measured flux densities. The rms of the scatter is $\sim$0.4 mas in each coordinate, much larger than the standard errors of the position determinations and most likely dominated by the fluctuation of the positions of the stellar radio emission relative to the center of the primary of \IMP.
Figure~\ref{f8IMschemradio} gives an artist's three-dimensional rendition of \IMP\ with the
primary as a giant with dark spots and the secondary as a sun-like star, each in its estimated orbit.

\subsection{Systematic and total errors}
A detailed analysis of systematic errors affecting each of the nine fit parameters of \IMP's motion relative to the distant universe led us to the following conclusions. For the position at the mean epoch of the VLBI observations of 2001.29, the systematic error in each coordinate was computed from 0.5 times the angular radius of the primary along the sky projection of the normal to the binary orbital plane (p.a.\ = 130.5\arcdeg\ 
$\pm$ 8.6\arcdeg). For the later epoch of 2005.08 in 
Table~\ref{tfinal}, the value for each component was added in quadrature with a possible maximum rms drift during that time interval of the radio brightness reference point of \IMP.
The rms drift rate was assumed to be not larger than
one stellar radius over the VLBI observation time span of 8.5 years or 0.06 \masyr\ in \RA\ and 0.05 \masyr\ in \DEC.

The systematic error of the proper motion estimate was found to be mostly comprised of the 
upper limit of the proper motion of C1 in the CRF and of the aforementioned possible maximum rms drift of the radio brightness reference point of \IMP. 

The systematic errors of the parallax and orbit estimates were found to be most clearly defined by the position values of B2250+194 relative to those of C1 in 3C 454.3, since the true parallax and orbit parameter values should be zero. The same parameter fit used for \IMP\ resulted in a parallax estimate for B2250+194 relative to C1 of $-0.032$ $\pm$ 0.074 mas, with the statistical standard error decreasing to 0.026 mas for $\chinu=1$. For the five-fold smaller separation of \IMP\ from 3C~454.3 the systematic uncertainty was estimated to be $<$0.015~mas in each coordinate. The same parameter fit also resulted in orbit parameter estimates for B2250+194 relative to C1 much smaller
than the estimates for \IMP\ itself.

For each of the parameters in Table~\ref{tfinal} the total error is also listed. In general it is computed as the root-sum-square of the statistical standard error and the estimated systematic error. For the position and proper motion values, however, the statistical errors are doubled before computing the root-sum-square to allow for correlated noise in the VLBI positions. For the parallax and orbit parameter estimates, the systematic errors were too small to contribute appreciably to the total standard errors. 

\begin{deluxetable}{l c c c c}
\tablecaption{Final \IMP\ parameter estimates\tablenotemark{a}
\label{tfinal}}
\tabletypesize{\tiny}
\tablehead{
\colhead{Parameter~~~~~~~~}  
  & \colhead{~~~~~~~~~Estimate~~~~~}  & \colhead{~~~~Stat. stand. error~~~~}  
  & \colhead{System. error\tablenotemark{b}}
  & \colhead{Total stand. error\tablenotemark{c}}  }
\startdata
\sidehead{Non-orbit parameters:}
$\alpha$ at epoch 2005.08\tablenotemark{d} (errors in mas) &  
\Ra{22}{53}{2}{258612}   & 0.12 & 0.33 & 0.40 \\
$\delta$ at epoch 2005.08\tablenotemark{d} (errors in mas) &  
\dec{16}{50}{28}{16005} & 0.13 & 0.29 & 0.39 \\
$\mua*$\tablenotemark{e} (\masyr) &  $-20.833$  & 0.026 & 0.073  & 0.090  \\
$\mud$ (\masyr) &  $-27.267$  & 0.030 & 0.074  & 0.095  \\
Parallax (mas) &  $\phantom{-}
                     10.370$  & 0.074 & $<$0.015 & 0.074 \\[3pt]
\sidehead{Linear model orbit parameters:\tablenotemark{f}}
\Asa\ (mas) &  $-0.59$  & 0.10 & $\ll$0.1 & 0.10 \\
\Asd\ (mas) &  $-0.66$  & 0.11 & $\ll$0.1 & 0.11 \\
\Aca\ (mas) &  $\phantom{-}0.15$  & 0.09 & $\ll$0.1 & 0.09 \\
\Acd\ (mas) &  $-0.23$  & 0.11 & $\ll$0.1 & 0.11 \\[3pt]
\sidehead{Alternative orbit parameters:\tablenotemark{g}}
Semimajor axis (mas) &  0.89  & 0.09 & $\ll$0.1 & 0.09 \\ 
Axial ratio\tablenotemark{h}  &  0.30  & 0.13 &$\ll$0.1 & 0.13 \\ 
P.A.\ of ascending node\tablenotemark{i} (deg) &  40.5  & 8.6 & $\ll$8 & 8.6 \\ 
$T_{\rm conj}$ (heliocentric JD)\tablenotemark{j} &  2450342.56
  & 0.44 & $\ll$0.4 & 0.44 \\[3pt]
\enddata
\tablenotetext{a}{Taken from \pfive.}
\tablenotetext{b}{See text.
}
\tablenotetext{c}{ 
The total error computed as the root-sum-square of the statistical standard error and the estimated systematic error, except for the position and proper motion parameters where we first doubled the standard errors before computing the root-sum-square. The upper bounds on the systematic
errors in the orbit terms apply to the mean orbit of the radio emission,
and not to the corresponding orbital terms for the stellar binary. For more information, see text.}
\tablenotetext{d}{
The position given is the estimated position 
of the center of mass of the \IMP\ binary
at epoch JD 2453403.0 (2005 Feb 1, $\sim$2005.08), 
the approximate midpoint of the \GPB\ science data.
Along with the proper motion, the position is specified in the (J2000.0) 
coordinate system and is closely tied to the
ICRF2 \citep{FeyGJ2009}.
}
\tablenotetext{e}{ 
$ \mu_{\alpha*} = \mu_{\alpha}\cos\delta$.
}
\tablenotetext{f}{
In the linear model, the orbital contribution to \IMP's position at time $T$ is  
\Asa~sin~$[2\pi(T - T_{\rm conj})/P]$ + \Aca~cos~$[2\pi(T - T_{\rm conj})/P]$ 
in $\alpha$ and 
\Asd~sin~$[2\pi(T - T_{\rm conj})/P]$ + \Acd~cos~$[2\pi(T - T_{\rm conj})/P]$ 
in $\delta$, where $P$ = 24.64877~d is the (fixed) orbital period and 
$T_{\rm conj}$ is the (fixed) time of conjunction, JD 2450342.905, adopted 
from \citet{Marsden+2005}.
}
\tablenotetext{g}{Computed by iteration to convergence so that the orbit on the sky given by the weighted least-squares estimates of the alternative orbit parameters is identical to that given by the weighted least-squares estimates of the linear orbit model parameters.}

\tablenotetext{h}{The ratio of the minor axis to the major axis of the 
 sky-projected orbit.
}
\tablenotetext{i}{See Figure~\ref{f8IMschemradio}
for illustration of the orbit geometry.  The
orbital motion on the sky is counterclockwise.
}
\tablenotetext{j}{Time of conjunction for the radio emitting region, that is 
for the conjunction nearest the one 
with the primary in back, i.e., at its greatest distance from us,
for the optical orbit of \protect\citet{Marsden+2005}.
}
\end{deluxetable}

\subsection{Comparison of our parameter estimates with previous estimates}
\label{comparison}
Comparing our results with the most precise previous measurements of proper motion and parallax, we find that our estimates agree with those in the {\em Hipparcos} Catalogue \citep{PerrymanESAshort1997}
and those of \citet{Lestrade+1999} within their respective larger standard errors. However, we found 
that our estimates slightly disagree with the values in the {\em Hipparcos} re-reduction \citep{vanLeeuwenF2007,vanLeeuwenF2008} by 1.6 and 2.4 times the combined standard error in \mud\ and in parallax, respectively. This discrepancy in \mud\ is, however, more than 10 times smaller than the standard error of either  the geodetic or the frame-dragging effect \citep{Everitt+2011}, and therefore of no consequence for the \GPB\ results.

Comparing the orbital parameters, we find that our estimate of the time of conjunction is consistent with that of \citet{Marsden+2005} and the combination of our estimates of orbital inclination and semimajor axis are consistent with their value of $a \sin i$. This agreement, in turn, is consistent with the reasonable expectation that the radio emission is on average centered on the primary of the \IMP\ system, and that it orbits with the same inclination. Any offset in phase of the radio orbit from that of the primary corresponds to less than one-forth of the radius of the primary.


\subsection{Erratic radio emission around the primary of \IMP}
\label{erratic}
\begin{figure}
\centering
\includegraphics[width=0.80\textwidth]{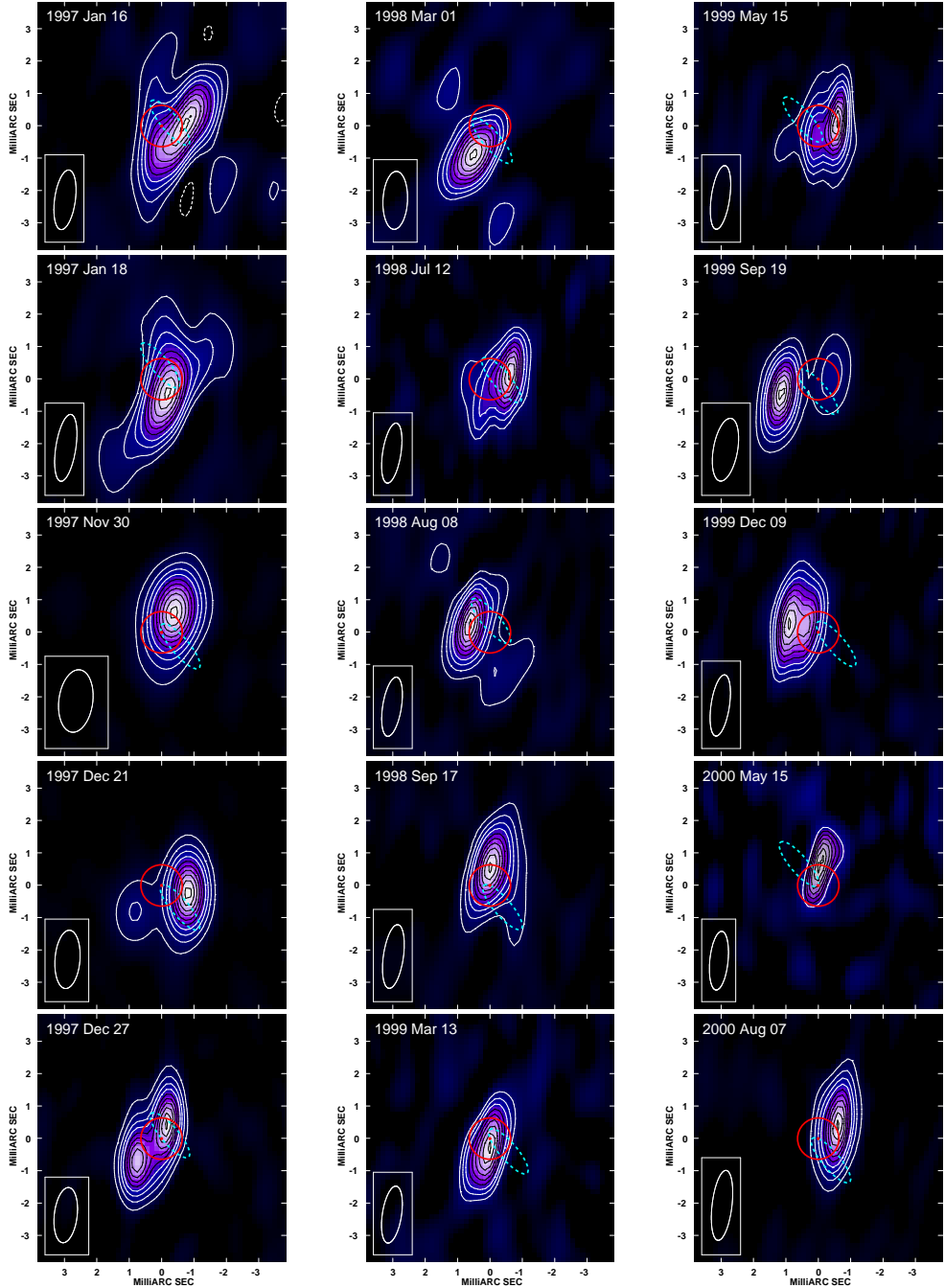}
\caption{A selection of VLBI images of \IMP\ at 8.4~GHz with their observing date and FWHM of the convolving beam.  North is up and east to
the left. Both the contours and the color scale show the brightness.
The contours are drawn at 10, 20, 30, \dots, 80, 90, and 98\% of the
peak brightness, starting with the first contour above $3\times$ the
rms background; and contours at 50\% and above are drawn in black.
The center of each panel (red dot) is the fit
position of the star's center, as derived from the astrometric
results in \pfive.  In other words, the coordinate origin
in our radio image should approximately represent the center of the
disk of
the primary star.  The red circle indicates the angular size of the
primary star
\citep[radius of $13.3 \pm 0.6$~\protect\Rsol;][]{BerdyuginaIT1999}.
The cyan dotted ellipse shows the orbit of the primary 
as in \pfive\ and \psix.  Figure taken from  \pseven, where also the complete set of our VLBI images
is given.
}
\label{f9image}
\end{figure}

With position at epoch, proper motion, parallax, orbital motion, and average location relative to the center of the primary determined, how can the radio emission be described and where does it
appear relative to the primary of \IMP\ from epoch to epoch? Figure~\ref{f9image} displays a set of VLBI images of \IMP\ relative to the center of the primary in its orbit about the binary's barycenter.

The analysis of the images showed that the radio emission is highly variable and frequently partly circularly polarized.  The morphology is also variable with the average size of the radio emission slightly larger than the disk of the primary.
The positions of the peaks of the emission regions are scattered over an area on the sky slightly larger than the disk of the primary.  The scatter has a tendency to be distributed along the orbit normal and expected primary's spin axis. 
Comparison with simulations suggest that the brightness peaks preferentially occur near the polar regions similar to the dark spots in the optical \citep{BerdyuginaM2006, Berdyugina+2000}. The height of the emission is likely close to 
the surface of the primary with 2/3 of the emission peaks located within 0.25 times the stellar radius above the surface.
The radio emission may be due to flares linked to a possibly dipolar magnetic field, whose axis is 
normal to the orbit plane, as could be expected giving the tidally locked rotation of the primary.  For a movie of the star based on these VLBI images, see the website, www.yorku.ca/bartel/impeg.mpg, or the online version of \pseven.

\section{Summary and conclusion}
\label{conc}
Our series of 35 VLBI sessions between 1997 and 2005 resulted in the most comprehensive and detailed
radio investigations ever made of a star.

\begin{itemize}

\item[1.] The proper motion of \IMP\ relative to the distant universe is $-20.83$  $\pm$ 0.09 \masyr\ in \al\ and $-27.27$ $\pm$  0.09 \masyr\ in \de, where the errors are intended to represent one-standard deviation errors and include estimates of systematic errors.
\item[2.] These results met the pre-launch requirements of the \GPB\ mission of standard errors no larger than 0.14 \masyr\ in each coordinate so as not to discernibly degrade the estimates of the geodetic and 
frame-dragging effects.
\item[3.] The parallax of \IMP\ is
10.37 $\pm$ 0.07~mas, corresponding to a distance of 96.4 $\pm$ 0.7~pc.
\item[4.] The proper-motion and parallax estimates agree with previous estimates, including those in the {\em Hipparcos} Catalogue, within the combined
standard errors, with the exception of the revised {\em Hipparcos} results of \citet{vanLeeuwenF2008} where the agreement is within 1.6 and 2.4 times the respective combined standard errors. Any discrepancy at these levels is however of no consequence for the \GPB\ results.
\item[5.] The radio emission of \IMP\ is highly variable.
\item[6.] The VLBI images of \IMP\ show emission regions slightly larger than the disk of the primary and their peaks scattered about a similarly sized area, consistent with being located preferentially near the polar regions and within $\sim$0.25 stellar radii above the surface.
\item[7.] The VLBI images of \IMP\ were assembled for a movie of the star (see the website of the first author, www.yorku.ca/bartel/impeg.mpg, or the online version of \pseven).
\end{itemize}

\acknowledgements 
ACKNOWLEDGEMENTS.  We thank each of the following
people, listed alphabetically, for their contributions of observations
and data reductions to our investigations of \IMP\ and its sky
neighborhood, for providing us with unpublished updated catalog data,
and/or for enlightening discussions on various aspects of this research
program:
S.~Berdyugina,
A.~Buffington, 
D.~Buzasi, 
N.~Caldwell, 
R.~Campbell,
J.~Chandler,
T.~Dame.
R.~Donahue, 
G.~Fazio, 
D.~Fischer, 
O.~Franz, 
G.~Gatewood,
D.~Gordon,
E.~Guinan, 
P.~Hemenway, 
A.~Henden,
M.~Holman, 
the late J.~Huchra, 
P.~Kalas, 
G.~Keiser, 
J.~Kolodziejczak, 
R.~Kurucz,  
B.~Lange, 
D.~Latham, 
J.~Lederman,
J.-F.~Lestrade,
J.~Lipa, 
P.~Luca,
S.~Marsden,
H.~McAlister, 
D.~Monet, 
N.~Nunes,
L.~Petrov,
R.~Schild, 
G.~Torres, 
W.~Traub, 
D.~Trilling, %
N.~Turner, 
the late H.~Wendker, 
and the late C.~Worley, 

This research was primarily supported by NASA, through a contract from
Stanford University to SAO, with a major subcontract from SAO to York
University.  We obtained observations made with the 100-m telescope of
the Max-Planck-Institut f\"{u}r Radioastronomie at Effelsberg,
with antennas of the National Radio Astronomy Observatory (NRAO),
which is a facility of the National Science Foundation operated under
cooperative agreement by Associated Universities, Inc., with antennas
of the DSN, operated by JPL/Caltech, under contract with NASA, and
with the NASA/ESA Hubble Space Telescope, which is operated by the
Association of Universities for Research in Astronomy, Inc., under
NASA contract NAS 5-26555. We made use of the SAO/NASA Astrophysics Data System
Abstract Service, initiated, developed, and maintained at SAO.

\end{document}